\documentclass[12pt]{article}

\usepackage{longtable}
\usepackage{amsmath}
\usepackage{amssymb}

\usepackage{graphicx}

\usepackage{cite}

\usepackage{color}

\usepackage{setspace} 
\usepackage{lineno}

\usepackage[labelfont=bf,labelsep=period,justification=raggedright]{caption}

\makeatletter
\renewcommand{\@biblabel}[1]{\quad#1.}
\makeatother

\date{}

 \newcommand{\startsupplement}{%
        \setcounter{table}{0}
        \renewcommand{\thetable}{S\arabic{table}}%
        \setcounter{figure}{0}
        \renewcommand{\thefigure}{S\arabic{figure}}%
 }

\begin{document}

\begin{flushleft}
{\Large
\textbf{Biases in the Experimental Annotations of Protein Function and their Effect on Our
Understanding of Protein Function Space}
}
\\
\author{Alexandra M. Schnoes$^1$%
       \email{Alexandra Schnoes - alexandra.schnoes@ucsf.edu}%
         David C. Ream$^2$%
         \email{David Ream - reamdc1@miamioh.edu}
      \and
         Alexander Thorman$^2$%
         \email{Alexander Thorman - thormanaaw@miamioh.edu}
      \and
         Patricia Babbitt$^1$%
         \email{Patricia Babbitt - babbittp@ucsf.edu}
       \and 
        Iddo Friedberg\correspondingauthor$^{2,3}$%
        \email{Iddo Friedberg\correspondingauthor - i.friedberg@miamioh.edu}
      }

\bf{1} Alexandra M. Schnoes Department of Bioengineering and Therapeutic Sciences,
University of California, San Francisco, San Francisco, CA,
USA
\\
\bf{2} David C. Ream Department of Microbiology, Miami University, Oxford, OH, USA
\\
\bf{3} Alexander W. Thorman, Department of Microbiology, Miami University, Oxford, OH, USA
\\
\bf{4} Patricia C. Babbitt Department of Bioengineering and Therapeutic Sciences, 
University of California, San Francisco, San Francisco, CA,
USA
\\
\bf{5} Iddo Friedberg, Departments of Microbiology and Computer Science \& Software engineering, 
Miami University, Oxford, OH, USA
\\
$\ast$ E-mail: corresponding i.friedberg@miamioh.edu
\end{flushleft}

\linenumbers 
\section*{Abstract}

The ongoing functional annotation of proteins relies upon the work of curators to capture
experimental findings from scientific literature and apply them to protein sequence and
structure data.  However, with the increasing use of high-throughput experimental assays,
a small number of experimental studies dominate the functional protein annotations
collected in databases.  Here we investigate just how prevalent is the ``few articles --
many proteins'' phenomenon.  We examine the experimentally validated annotation of proteins 
provided by several
groups in the GO Consortium, and show that the distribution of proteins per published study
is exponential, with 0.14\% of articles providing the source of annotations for 25\% of the
proteins in the UniProt-GOA compilation. Since each of the dominant articles describes the
use of an assay that can find only one function or a small group of functions, this leads
to substantial biases in what we know about the function of many proteins.
Mass-spectrometry, microscopy and RNAi experiments dominate high throughput experiments.
Consequently, the functional information derived from these experiments is mostly of the
subcellular location of proteins, and of the participation of proteins in embryonic developmental
pathways. For some organisms, the information provided by different studies overlap by a
large amount. We also show that the information provided by high throughput experiments is
less specific than those provided by low throughput experiments. Given the experimental
techniques available, certain biases in protein function annotation due to high-throughput
experiments are unavoidable. Knowing that these biases exist and understanding their
characteristics and extent is important for database curators, developers of function
annotation programs, and anyone who uses protein function annotation data to plan
experiments.

\section*{Author Summary}

Experiments and observations are the vehicles used by science to understand the world around
us. In the field of molecular biology, we are increasingly relying on high-throughput, genome-wide
experiments to provide answers about the function of biological macromolecules.  However, any
experimental assay is essentially limited in the type of information it can discover. Here we show
that our increasing reliance on high-throughput experiments biases our understanding of protein
function. While the primary source of information is experiments,  the functions of many proteins are
computationally annotated by sequence-based similarity, either directly or indirectly, to proteins whose
function is experimentally determined. Therefore, any biases in experimental annotations can get
amplified and entrenched in the majority of protein databases. We show here that high-throughput
studies are biased towards certain aspects of protein function, and that they provide less
information than low-throughput studies. While there is no clear solution to the phenomenon of bias
from high-throughput experiments, recognizing its existence and its impact can help take steps to
mitigate its effect.

\section*{Introduction}

Functional annotation of proteins is an open problem and a primary challenge in molecular
biology today\cite{Friedberg2006Automated, Schnoes2009Annotation, Erdin2011180,
Rentzsch2009210}. The ongoing improvements in sequencing technology have shifted the emphasis
shifting from realizing the \$1,000 genome to realizing the 1-hour genome
\cite{Stahl2012Toward}. The ability to rapidly and cheaply sequence genomes is creating a flood
of sequence data, but to make these data useful, extensive analysis is needed. A large portion
of this analysis involves assigning biological function to newly determined gene sequences, a
process that is both complex and costly\cite{Sboner2011Real}.  To aid current annotation
procedures and improve computational function prediction algorithms, high-quality and
experimentally derived data are necessary. Currently, one of the few repositories of
such data is the UniProt-GOA database\cite{Dimmer2012UniProtGO}, which is a compilation of data
contributed by several member groups of the GO consortium.  UniProt-GOA contains functional
information derived from literature, and by computational means. The information derived from
literature is extracted by human curators who capture functional data from publications, assign
the data to their appropriate place in the Gene Ontology hierarchy\cite{Ashburner2000Gene} and
label them with appropriate functional evidence codes.  
UniProt-GOA is compiled from annotations made
by several member groups of the GO consortium, and as such presents the  current state of our
view of protein function space.  It is therefore important to understand any trends and biases
that are encapsulated in UniProt-GOA, as those impact well-used sister databases and
consequently a
large number of users worldwide. 

One concern surrounding the capture of functional data from articles is the propensity for
high-throughput experimental work to become a large fraction of the data in the GO Consortium database,
thus having  a small number of experiments dominate the protein function landscape.  In
this work we analyzed the relative contribution of peer-reviewed articles describing all
the experimentally derived annotations in UniProt-GOA. We found some striking trends,
stemming from the fact that a small fraction of articles describing high-throughput
experiments disproportionately contribute to the pool of experimental annotations of model
organisms. Consequently we show that: 1) annotations coming from high-throughput
experiments are overall  less informative than those provided by low-throughput
experiments;  2) annotations from high-throughput experiments are biased towards a limited number of
functions, and, 3) many high-throughput experiments overlap in the proteins they annotate,
and in the annotations assigned. Taken together, our findings offer a  picture of how the
protein function annotation landscape is generated from scientific literature.
Furthermore, due to the biases inherent in the current system of sequence annotations, this
study serves as a caution to the producers and consumers of biological data from
high-throughput experiments.

\section*{Results}

\subsection*{Articles and Proteins} 

\begin{figure}[!ht]
\begin{center}
\includegraphics[width=6in]{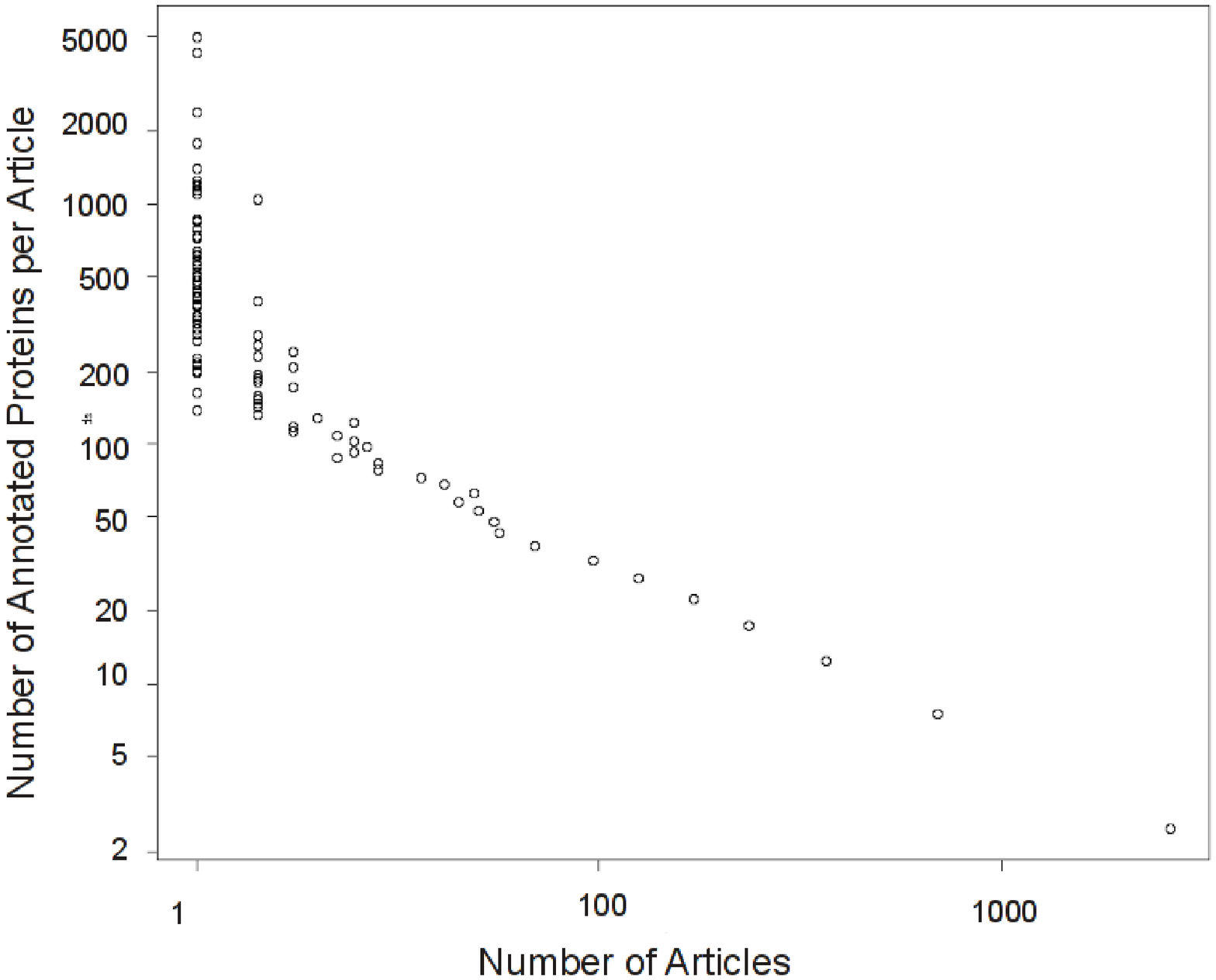}
\end{center}
\caption{
{\bf Distribution of the number of proteins annotated per article.} X-axis: number of annotating
articles. Y-axis: number of annotated proteins. The distribution was found to be logarithmic with a
significant ($R^2=0.72; p<1.10\times 10^{-18}$) linear fit to the log-log plot. The data came from 76137
articles annotating 256033 proteins with GO experimental evidence codes, in Uniprot-GOA 12/2011.
}
\label{fig:articles-prots}
\end{figure}

The increase in the number of high-throughput experiments used to determine protein functions may
introduce biases into experimental protein annotations, due to the inherent capabilities and limitations
of high-throughput assays. To test the hypothesis that such biases exist, and to study their extent if
they do, we compiled the details of all experimentally annotated proteins in UniProt-GOA.  This included
all proteins whose GO annotations have the GO experimental evidence codes EXP, IDA, IPI, IMP, IGI, IEP
(See Methods for an explanation of GO evidence codes).  We first examined the distribution of articles
that are the source of experimentally validated annotations by the number of proteins they annotate.  As
can be seen in Figure~\ref{fig:articles-prots}, the distribution of the number of proteins annotated per
article follows a power-law distribution. $f(x)=ax^k$.  Using linear regression over the log values of the
axes we obtained a fit with $p<1.18\times 10^{-8}$ and $R^2=-0.72$.  We therefore conclude that there is
indeed a substantial bias in experimental annotations, in which there are few articles that annotate a
large number of proteins.

To better understand the consequences of such a distribution, we divided the
annotating articles into four cohorts, based on the number of proteins each article
annotates.  \textit{Single-throughput} articles are those articles that annotate
only one protein; \textit{low throughput} articles annotate 2-9 proteins;
\textit{moderate throughput} articles annotate 10-99 proteins and \textit{high
throughput} articles annotate over 99 proteins. The results are shown in
Table~\ref{tab:cohorts}. The most striking finding is that high throughput articles
are responsible for 25\% of the annotations that the GO Consortium creates, even though they
are found only in 0.14\% of the articles. 96\% of the articles are single-throughput and 
low-throughput, however those annotate only 53\% of the proteins. So
while moderate-throughput and high-throughput studies account for almost 47\% of
the annotations in Uniprot-GOA, they constitute only 3.66\% of the studies published.

To understand how the log-odds distribution affects our understanding of protein function,
we examined different aspects of the annotations in the four article cohorts. Also, we
examined in greater detail the top-50 high-throughput annotating articles.  ``Top-50 high
throughput annotating articles'' are the articles describing experimental annotations
that are top ranked by the number of proteins annotated per article. An initial
characterization of these articles is shown in Table~\ref{tab:top50}. As can be seen, most
of the articles are specific to a single species (typically a model organism) and to a
single assaying pipeline that is used to assign function to the proteins in that organism.
With one exception, only one ontology of the three GO ontologies was used for annotation in
any single experiment. The three ontologies are Molecular Function (MF), Biological Process
(BP) and Cellular Component (CC).  These are separate ontologies within GO, describing
different aspects of function as detailed in\cite{Ashburner2000Gene}.  As we show later,
for some species this means that a single functional aspect (MF, BP or CC) of a species can
be dominated by a single study.

\subsection*{The Impact of High Throughput Studies on the Annotation of Model Organisms}

\begin{figure}[!ht]
\begin{center}
\includegraphics[width=6in]{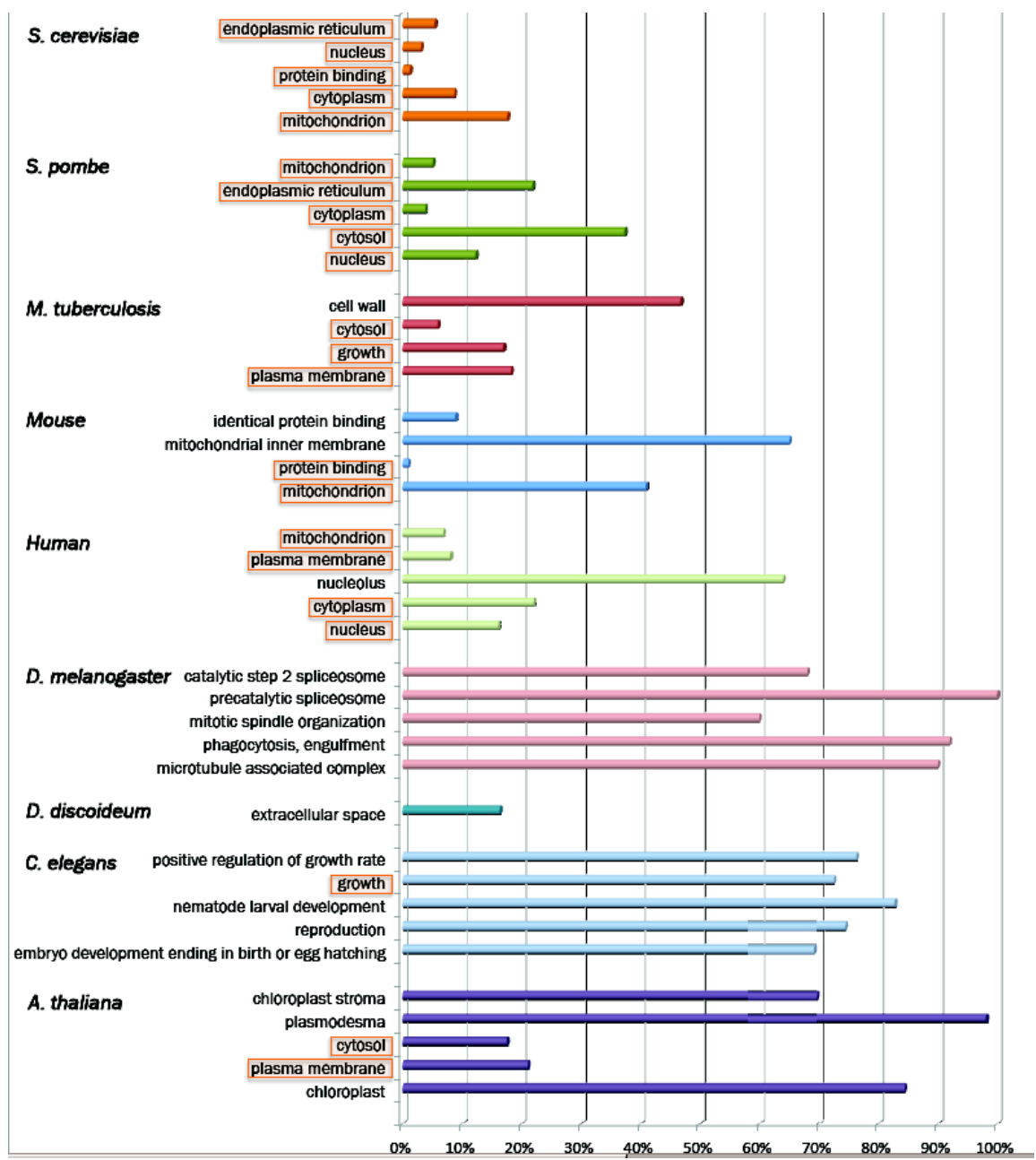}
\end{center}
\caption{
{\bf Relative contribution of top-50 articles to the annotation of major model organisms.}  
The length of each bar represents the percentage of proteins annotated by the top-50 articles in a
given organism by a given GO term. GO terms that are present in more than one species are
highlighted. 
}
\label{fig:rel-contrib}
\end{figure}

We examined the relative contribution of the
top-50 articles to the entire corpus of experimentally annotated proteins in each
species.  Unsurprisingly, all the species found in the top-50 articles were either
common model organisms or human.  For each species, we examined the five most
frequent terms in the top-50 articles. We then examined the contribution
of this term by the top-50 articles to the general annotations of that species.  The
\textit{contribution} is the number of annotations by any given GO term in the top
50 articles divided by the number of annotations by that GO term in all of
UniProt-GOA.  For example, as seen in Figure~\ref{fig:rel-contrib} in \textit{D.
melanogaster} 88\% of the annotations using the term ``precatalytic splicosome'' in
articles experimentally annotating this species are contributed by the top-50
articles. 


For most organisms annotated by the top-50 articles, the annotations were within the cellular
component or biological process ontologies. Notable exceptions are \textit{D. melanogaster} and \textit{C.
elegans} where the dominant terms were from the Biological Process ontology, and in
mouse, where ``protein binding'' and ``identical protein binding'' are from the
Molecular Function Ontology.  \textit{D. melanogaster}'s annotation for the top
terms is dominated (over 50\% contribution) by the top-50 articles. 

The term frequency bias described here can be viewed more broadly within the
ontology bias. The proteins annotated by the cohorts of single-protein articles,
low-throughput articles, and moderate-throughput articles have similar ratios of the
fraction of proteins annotated.  Twenty-two to twenty-six percent of assigned terms
are in the Molecular Function Ontology, and 51-57\% are in the Biological Process
Ontology and the remaining 17-25\% are in the Cellular Component ontology.  These
ratios change dramatically with high-throughput articles (over 99 terms per
article). In the high-throughput articles, only 5\% of assigned terms are in the
Molecular Function Ontology, 38\% in the Biological Process Ontology and 57\% in the
Cellular Compartment Ontology, ostensibly due to a lack of high-throughput assays
that can be used for generating annotations using the Molecular Function Ontology. 

\subsection*{Repetition and Consistency in Top-50 Annotations}

\begin{figure}[!ht] 
\begin{center} 
\includegraphics[width=7in]{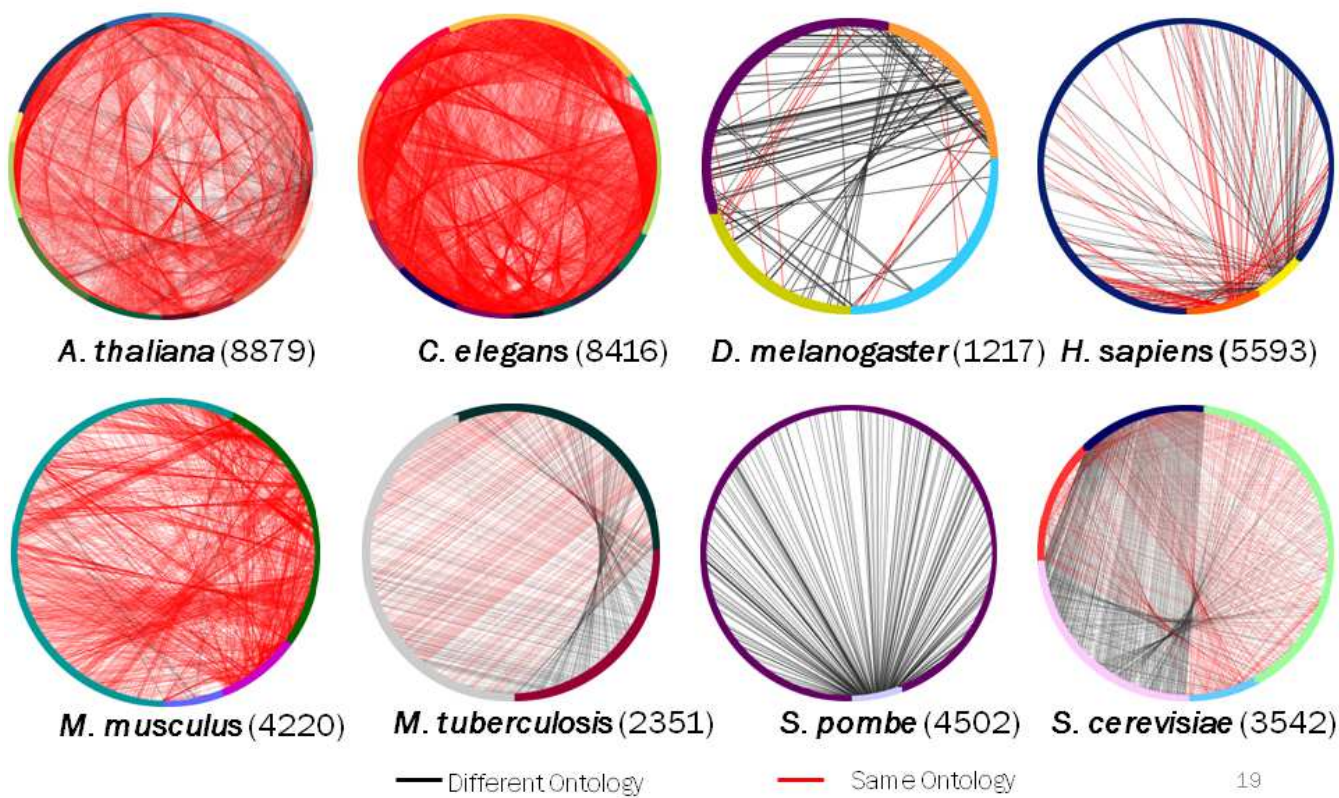} 
\end{center}
\caption{\textbf{Redundancy in proteins described by the top-50 articles.} A circle represents the sum
total of articles annotating each organism. Each colored arch is composed of all the proteins in a
single article. A line is drawn between any two points on the circle if the proteins they represent
have 100\% sequence identity. A black line is drawn if they are annotated with a different ontology
(for example, in one article the protein is annotated with the MFO, and in another article with BPO); a red
line if they are annotated in the same ontology. Example: \textit{S. pombe} is described by two
articles, one with few protein (light arch on bottom) and one with many (dark arch encompassing most
of circle). Many of the same proteins are annotated by both articles. See Table~\ref{tab:dreamcatcher1}
for numbers.} 
\label{fig:dreamcatcher1} 
\end{figure}

How many of the top-50
articles actually annotate the same set of proteins? Answering this question will tell us how
repetitive experiments are in identifying the same set of proteins to annotate. However, even
when annotating the same set of proteins and within the same ontology, different experiments may
provide different results, lacking consistency. Therefore, the annotation consistency was also checked.
Repetition is given as $n\over N$ with $n$ being the number of proteins annotated by two or more articles, 
and $N$ being the total number of proteins.
The results of the repetition analysis are shown in
Figure~\ref{fig:dreamcatcher1} and in Table~\ref{tab:dreamcatcher1}. As can be seen, the
highest repetition (65\%) is in the 12 articles annotating \textit{C. elegans}. Of course,
a higher number of articles is expected to increase repetitive annotations simply due to
increased sampling of the genome.  However, the goal of this analysis is to present the
degree of repetition, rather than to try to rank and normalize it. As an additional
repetition metric, Table~\ref{tab:dreamcatcher1}  also lists the mean number of sequences
per cluster.
When normalized by number of annotating articles, the highest repetition
is found in Mouse (15.33\% in three articles) closely followed by \textit{M.
tuberculosis} (14\% in two articles). Taken together, these results show that
there is repetition in choosing the proteins that are to be annotated in most model organisms using
high-throughput assays, although the rate of this repetition varies widely.

Consistency analysis took place as described in Methods. The consistency measure
is normalized on a 0-1 scale, with 1 being most consistent, meaning that all annotations from all
sources are identical.  Table~\ref{tab:dreamcatcher2} shows the results of this analysis. In
\textit{A. thaliana}, 1941 proteins are annotated by 15 articles and 18 terms in the Cellular
Component ontology.  The mean maximum-consistency is 0.251. The highest mean consistency is for
the annotation of 807 mouse proteins annotated in Cellular Component ontology with an annotation
consistency 0.832. However, that is not surprising given that there are only three annotating
articles, and two annotating terms. We omitted the ontology and organism combinations that were
annotated by less than three articles or two GO terms, or both.

\subsection*{Quantifying Annotation Information}

A common assumption holds that while high-throughput experiments do annotate more protein
functions than low-throughput experiments, the former also tend to be more shallow in the
predictions they provide. The information provided, for example, by a large-scale protein binding
assay will only tell us if two proteins are binding, but will not reveal whether that binding is
specific, will not provide an exact $K_{bind}$, will not say under what conditions binding takes
place, or whether there is any enzymatic reaction or signal-transduction involved. Having on hand
data from experiments with different ``throughputness'' levels,  we set out to investigate
whether there is indeed a difference in the information provided by high-throughput experiments vs.
low-throughput ones. We examined the information provided by GO terms in each paper cohort using two
methods: edge-count, and information-content. See Methods for details.

\begin{figure}[!ht]
\begin{center}
\includegraphics[width=6.0in]{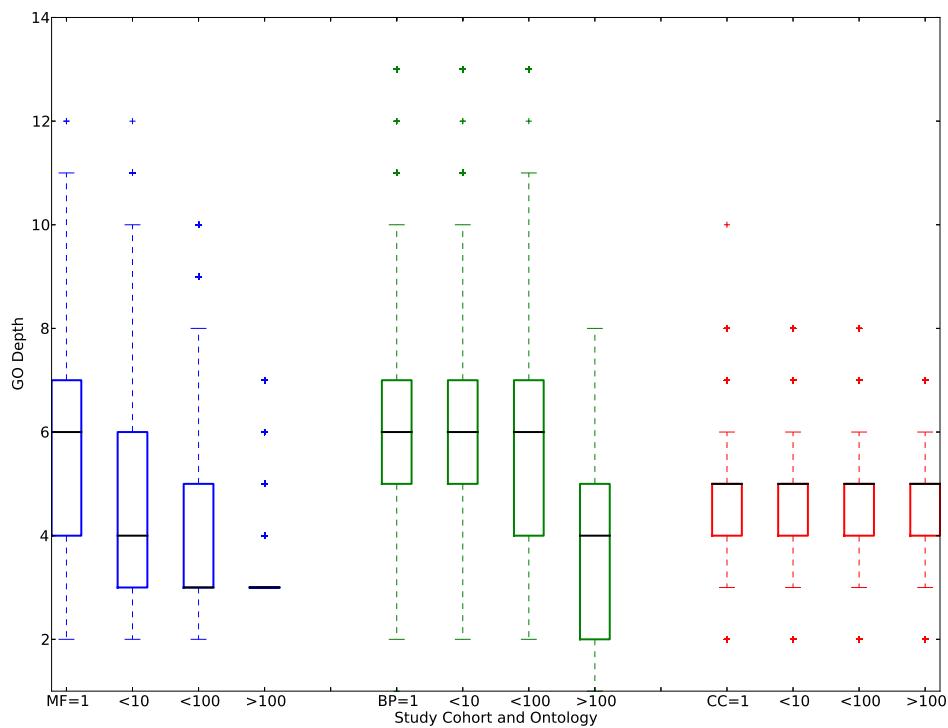}
\end{center}
\caption{
\textbf{Information provided by articles depending on the number of proteins the articles
annotate.} Articles are grouped into cohorts: 1: one protein annotated by article; $<10$:
more than 1, up to 10 annotated; $<100$: more than 10, less than 100 annotated; $\geq100$:
100 or more proteins annotated per article. Blue bars: Molecular Function ontology; Green
bars: Biological Process ontology; Red bars: Cellular Component ontology. Information is
gauged by {\bf A}: Information Content and {\bf B}: GO depth.  See text for details.}
\label{fig:go-depth}
\end{figure}

The results of both analyses are shown in Figure~\ref{fig:go-depth}. 
In general, the results from the edge count analysis and the information-content based analysis are in
agreement when compared across annotation cohorts.
For the Molecular Function ontology, the distribution of edge counts and log-frequency
scores decreases as the number of annotated proteins per-article increases.  For the Biological
Process ontology, the decrease is significant. However the contributors to the decrease are the
high-throughput articles while there is little change in the first three article cohorts.
Finally, there is no significant trend of GO-depth decrease in the Cellular Component Ontology.
However, using the information-content metric, there is also a significant decrease in
information-content in the high-throughput article cohort.

\subsection*{Exclusive High Throughput Annotations}

Of interest is the fraction of proteins that are exclusively annotated by high-throughput
experiments. The question here is: from the experimentally annotated proteins in an
organism, how much do we know of their function \textit{only} using high-throughput
experiments?  We have seen that high-throughput experiments annotate a large number of
proteins, but still some 80\% of experimentally determined proteins are annotated via
medium-, low- and single-throughput experiments. Given the lower information-content of
high-throughput experiments, it is important to know which organisms have a substantial
fraction of the proteins experimentally annotated by high throughput studies only. To do so,
we analyzed all species with more than 200 genes in the NCBI taxa database for the fraction
of the genes that are  exclusively annotated  by high throughput studies. The results are
shown in Table~\ref{tab:exclusive_ht}.

As can be seen, although the fraction of high-throughput annotated proteins is large, 
not many species are affected with a large fraction of proteins that are exclusively annotated
by high-throughput studies. However, the few species that are affected are important study
and model species.  It is important to note that some redundancy due to isoforms, mutants and
duplications may exist.

\subsection*{Frequently Used High-Throughput Experiments}

The twenty GO evidence codes, discussed above, encapsulate the means by which the function was
inferred, but they do not capture all the necessary information. For example, ``Inferred by
Direct Assay'' (IDA) informs that some experimental assay was used, but does not say which type
of assay.  This information is often needed, since knowing which experiments were performed can
help the researcher establish the reliability and scope of the produced data. RNA, used in an
RNAi experiment does not traverse the blood-brain-barrier, meaning that no data from the central
nervous system can be drawn from an RNAi experiment. The Evidence Code Ontology, or ECO, seeks
to improve upon the GO-attached evidence codes. ECO provides more elaborate terms than
``Inferred by Direct Assay'': ECO also conveys which assay was used, for example ``microscopy''
or ``RNA interference''.  In addition to evidence terms, the ECO ontology provides
\textit{assertion terms} in which the nature of the assay is given. For example, an
enzyme-linked immunosorbent assay (ELISA) provides quantitative protein data \textit{in vitro}
while an immunogold assay may provide the same information, and cellular localization
information \textit{in situ}. We manually assigned Evidence Codes Ontology (ECO) assertion and
evidence terms to the top-50 articles.  
The assignment is shown in detail in Table~\ref{tab:sup:papers_eco}. Table~\ref{tab:evidence}
shows the sorted count of ECO terms in the top-50 papers.

The most frequent ECO term used is ECO:0000160 ``protein separation followed by fragment identification
evidence'': this fits the 27 papers that essentially describe mass-spectrometry studies.
Consequently this means that the assignment procedure is limited to the cellular
compartments that can be identified with the fractionation methods used. So while Cellular
Component is the most frequent annotation used, fractionation and mass-spectrometry is the
most common method used to localize proteins in subcellular compartments. A notable
exception to the use of fractionation and MS for protein localization is in the top
annotating article \cite{pmid18029348}, which uses microscopy for subcellular localization.

The second most frequent experimental ECO term is ``Imaging assay evidence'' (ECO:000044).  Several types
of studies fall under this ECO. Those include microscopy, RNAi, some of the mass-spectrometry studies that
used microscopy, and a yeast-2-hybrid study. As imaging information is used in a variety of studies, this
ECO term is not informative of the chief method used in any study, but rather the importance of imaging
assays in a variety of methods.  The third most frequent experimental ECO term used was ``Cell
fractionation evidence'' which is closely associated with the top term, ``Imaging assay evidence''.  The
fourth annd fifth most frequent ECO term used were ``loss-of-function mutant phenotype evidence''
(ECO:0000016) and ``RNAi evidence'' (ECO:000019). These two terms are also closely associated, in RNAi
whole-genome gene knockdowns in \textit{C. elegans}, \textit{D.  melanogaster} and one in \textit{C.
albicans}. RNAi experiments use targeted dsRNA which is delivered to the organism and silences specific
genes. Typically the experiments here used libraries of RNAi targeted to the whole exome (for example
\cite{pmid11099033,pmid11231151, pmid12445391,pmid12529635}). The phenotypes searched for were mostly
associated with embryonic and post-embryonic development.  Some studies focused on mitotic spindle
assembly\cite{pmid17412918}, lipid storage\cite{pmid12529643} and endocytic traffic\cite{pmid17704769}.
One study used RNAi to identify mitochondrial protein localization \cite{pmid18433294}. These studies
mostly use the same RNAi libraries, and target the whole \textit{C.  elegans} genome using common data
resources. Hence the large redundancy observed for \textit{C. elegans} in Table~\ref{tab:dreamcatcher1}.
It should be noted that all experiments are associated with computational ECO terms, which
describe sequence similarity and motif recognition techniques used to identify the sequences found:
``sequence similarity evidence'', ``transmembrane domain prediction evidence'', ``protein BLAST evidence''
etc.  These terms are all bolded in Table~\ref{tab:evidence}.  A strong reliance on computational
annotation is therefore an integral part of high throughput experiments. It should be noted that
computational annotation here is not used directly for functional annotation, but rather for identifying
the protein by a sequence or motif similarity search.
The third most frequently used assertion in the top experimental articles was
not an experimental assertion, but rather a computational one: the term ECO:00053
``computational combinatorial evidence'' is defined as ``A type of combinatorial analysis
where data are combined and evaluated by an algorithm.'' This is not a computational
prediction \textit{per se}, but rather a combination of several experimental lines of
evidence used in a article.


\section*{Discussion}

We have identified several annotation biases in GO annotations provided by the GO consortium. These
biases stem from the uneven number of annotations produced by different types of experiments. It is
clear that results from high-throughput experiments contribute substantially to the function annotation
landscape, as up to 20\% of experimentally annotated proteins are annotated by high-throughput assays.
At the same time, high throughput experiments produce less information per protein than
moderate--, low-- and single-- throughput experiments as evidenced by the type of GO terms
produced in the Molecular Function and Biological Process ontologies. Furthermore, the
number of total GO terms used in the high-throughput experiments is much lower than that
used in low and medium throughput experiments. Therefore, while high throughput experiments
provide a high coverage of protein function space, it is the low throughput
experiments that provide more specific information, as well as a larger diversity of terms.

We have also identified several types of biases that are contributed by high throughput
experiments.  First, there is the enrichment of low-information-content GO terms, which means
that our understanding of the protein function as provided by high-throughput experiments is more
limited than that provided by low-throughput experiments.  Second, there is the small number of
terms used, when considering the large number of proteins that are being annotated. Third is the
general ontology bias towards the cellular component ontology and, to a lesser extent, the
Biological Process ontology: there are very few articles that deal with the
Molecular Function ontology.  These biases all stem from the inherent capabilities and
limitations of the hight-throughput experiments. A fourth, related bias is the organism studied:
taken together, studies of \textit{C. elegans} and \textit{A. thaliana} studies comprise 36 of
the top-50 annotating articles, or 72\%.

\subsection*{Information Capture and Scope of GO}

We have discussed the information loss that is characteristic of high-throughput experiments, as
shown in Figure~\ref{fig:go-depth}. However, another reason for information loss is the
inability to capture certain types of information using the Gene Ontology. GO is purposefully
limited to three aspects (MF, BP and CC) of biological function, which are assigned per protein.
However, other aspects of function may emerge from experiments. Of
note is the study, ``Proteome survey reveals modularity of the yeast cell
machinery''\cite{pmid18029348}. In this study, the information produced was primarily of protein
complexes, and the relationship to cellular compartmentalization and biological networks. At the
same time, the only GO term captured in the curation of proteins from this study was ``protein binding''.
Some, but not all of this information can be captured more specifically using the children of
the term ``protein binding'', but such a process is arguably laborious by manual curation of the
information from a
high throughput article.  Furthermore, the main information conveyed by this article, namely the
types of protein complexes discovered and how they relate to cellular networks, is outside the
scope of GO. It is important to realize that while high-throughput experiments do convey less
information per protein within the functional scope as defined by GO, they still convey
composite information such as possible pathway mappings -- information which needs to be
captured into annotation databases by means other than GO.  In the example above, the
information can be captured by a protein interaction database, but not by GO terms. Methods such
as the Statistical Tracking of Ontological Phrases\cite{Wittkop2013STOP} can help in selecting
the appropriate ontology for better information capture.

\subsection*{Conclusions}

Taken together, the annotation trends in high-throughput studies affect our understanding of
protein function space. This, in turn, affects our ability to properly understand the
connection between predictors of protein function and the actual function -- the hallmark of
computational function annotation. As a dramatic example, during the 2011 Critical Assessment of
Function Annotation experiment\cite{CAFA2013}  it was noticed that
roughly 20\% of the proteins participating in the challenge and annotated with the Molecular
Function Ontology were annotated as ``protein binding'', a GO term that conveys little
information. Furthermore, it was shown that the major contribution of ``protein binding''
term to the CAFA challenge data set was due to high-throughput assays. This illustrates how
the concentration of a large number of annotations in a small number of studies provides
only a partial picture of the function of these proteins. As we have seen, the picture
provided from high throughput experiments is mainly of: 1) subcellular localization cell
fractionation and MS based localization and 2) developmental phenotypes. While these data
are important, we should be mindful of this bias when examining protein function in the
database, even those annotations deemed to be of high quality, those with experimental
verification. Furthermore, such a large bias in prior probabilities can adversely affect
programs employing prior probabilities, as most machine-learning programs do. 
If the training set for these programs has included a disproportional number of
annotations by high-throughput experiments, the results these programs provide will be
strongly biased towards a few frequent and shallow GO terms.

To remedy the bias created by high throughput annotations, the provenance of annotations should
be described in more detail by curators and curation software. Many function annotation
algorithms rely on homology transfer as part of their pipeline to annotate query
sequences\cite{Friedberg2006Automated, CAFA2013}.  Knowing the annotation provenance, including
the number of proteins annotated by the original paper can create less biased benchmarks or
otherwise incorporate that information into the annotation procedure.  The ECO ontology can be
used to determine the source of the annotation, and the user or the algorithm can decide whether
to rely upon any combinations of ``throughputness'' and experimental type.  Of course, such
approaches should be taken cautiously, as sweeping measures can cause the unintended loss of
information. We hereby call upon the communities of annotators, computational biologists and
experimental biologists to be mindful of the phenomenon of the experimental biases described in
this study, and to work to understand its implications and impact.

\section*{Methods}
We used the UniProt-GOA database from December 2011.
Data analyses were performed using Python scripts. The following tools were used in the 
analyses: Biopython\cite{Cock2009Biopython}, matplotlib\cite{Hunter:2007}. 
ECO terms classifying the proteins in the top 50 experiments were assigned to the proteins
manually after reading the articles. 
All data and scripts are available on:
http://github.com/idoerg/Uniprot-Bias/ and on http://datadryad.org (the
latter upon acceptance). 

\subsection*{Use of GO evidence codes}
Proteins in UniProt-GOA are annotated with one or more GO terms
using a procedure described in Dimmer \textit{et al.} (2012). Briefly, this procedure consists
of six steps which include sequence curation, sequence motif analyses, literature-based
curation, reciprocal BLAST\cite{Altschul1997Gapped} searches, attribution of all resources
leading to the included findings, and quality assurance. If the annotation source is a research
article, the attribution includes its PubMed ID. For each GO term associated with a protein,
there is also an \textit{evidence code} which the curator assigns to explain how the association
between the protein and the GO term was made.  Experimental evidence codes include such terms
as: \textit{Inferred by Direct Assay} (IDA) which indicates that ``a direct assay was carried
out to determine the function, process, or component indicated by the GO term'' or
\textit{Inferred from Physical Interaction} (IPI) which ``Covers physical interactions between
the gene product of interest and another molecule.'' (All GO evidence code definitions were
taken from the GO site, geneontology.org.) Computational evidence codes include terms such as
\textit{Inferred from Sequence or Structural Similarity} (ISS) and \textit{Inferred from
Sequence Orthology} (ISO).  Although the evidence in computational evidence codes is
non-experimental, the proteins annotated with these evidence codes are still assigned by a
curator, rendering a degree of human oversight. Finally, there are also computational,
non-experimental evidence codes, the most prevalent being \textit{Inferred from Electronic
Annotation} (IEA) which is ``used for annotations that depend directly on computation or
automated transfer of annotations from a database''. IEA evidence means that the annotation is
electronic,  and was not made or checked by a person. Different degrees of reliability are
associated with different evidence codes, with experimental codes generally considered to be of
higher reliability than non-experimental codes. (For details see:
http://www.ebi.ac.uk/GOA/ElectronicAnnotationMethods)

\subsection*{Quantifying GO-term Information}
We used two methods to quantify the information given by GO terms.  First we used edge counting ,
where the information contained in a term is dependent on the edge distance of that term from the
root.  The term ``catalytic activity''(one edge distance from the ontology root node) would be less
informative than ``hydrolase activity'' (two edges) and the latter will be less informative than
``haloalkane dehalogenase activity'' (five edges).  We therefore counted edges from the ontology root
term to the GO term to determine term information. The larger the number of edges, the more specific
--and therefore informative-- is the annotation. In cases where several paths lead from the root to
the examined GO term, we used the minimal path. We did so for all the annotating articles split into
groups by the number of proteins each article annotates. 

While edge counting provides a measure of term-specificity, this measure is imperfect. The reason is
that each of the three GO ontologies is constructed as a directed acyclic graph (DAG) where different
areas of the GO DAG have different connectivities, and terms may have different depths unrelated to the
intuitive specificity of a term. For example ``D-glucose transmembrane transporter activity'',
(GO:0055056) is 10 terms deep, while ``L-tryptophan transmembrane transporter activity'', (GO:0015196)
is fourteen terms deep. It is hard to discern whether these differences are meaningful.  For this
reason, information content, the logarithm of the inverse of the GO term frequency in the corpus, is
generally accepted as a measure of GO term information
content\cite{Lord2003Investigating,Pesquita2009Semantic}. To account for the possible bias created by
the GO-DAG structure, we also used the log-frequency of the terms in the experimentally annotated
proteins in Uniprot-GOA.  However, it should be noted that the log-frequency measure is also imperfect
because, as we see throughout this study, a GO term's frequency may be heavily influenced  by the top
annotating articles, injecting a circularity problem into the use of this metric.  Since no single
metric for measuring the information conveyed by a GO term is wholly satisfactory, we used both
edge-counting and information-content in this study.

\subsection*{Annotation Consistency}
To examine annotation consistency, we employed the following method:
given a protein $P$, let $G$ be the terminal (leaf) GO terms $g_1,g_2,\dots,g_m$ that annotate that protein
in all top-50 articles for a single ontology $O \in \{BPO,MFO,CCO\}$. The count of each of these
GO terms per protein per ontology is $n_1,n_2,\dots,n_m$ with $n_i$ being the number of
times GO term $g_i$ annotates protein $P$.

The number of total annotations for a protein in an ontology is $\sum_1^m n_i$ . The
\textit{maximum annotation consistency} for protein $P$ in ontology $O$ $0 \leq k_{P,O}
\leq 1$ is calculated as:

\[
k_{P,O} = \frac{max(n_1,n_2,\dots,n_m)}{\sum^m n_i}\;  for \: max(n_1,n_2,\dots,n_m) \geq 2
\]

For example, the protein ``Oleate activated transcription factor 3'' (UniProtID: P36023) in
\textit{S.  cerevisiae} is annotated four times by three articles using the Cellular
Component ontology:

\begin{table}[!ht]
\begin{tabular}{l l l l l}
PubMedID  &  UniProt ID& Ontology & GO term & description \\
14562095  &  P36023    & CCO & GO:0005634  &   nucleus \\
14562095  &  P36023    & CCO  &  GO:0005737 &  cytoplasm \\
16823961  &  P36023    & CCO  & GO:0005739  &    mitochondrion \\
14576278  &  P36023    & CCO  & GO:0005739  &    mitochondrion \\
\end{tabular}
\end{table}

The annotation consistency for P36023 is therefore the maximum count of identical GO terms
(\textit{mitochondrion}, 2), divided by the total number of annotations, 4: 0.5.

When choosing a measure for annotation consistency, we favored a simple and interpretable measure. We therefore
examined identity among leaf terms only, rather than use a more complex comparison of multiple subgraphs in the GO
ontology DAG (Directed Acyclic Graph). Doing so without manual curation is unreliable, and may skew the perception of
similarity\cite{Faria2012Mining}.  

\section*{Acknowledgments}
We thank Predrag Radivojac, Nives \v{S}kunca, Cristophe Dessimoz, Yanay Ofran, Marc
Robinson-Rechavi, Wyatt Clark, Tony Sawford, Chris Mungall, Rama Balakrishnan,
and members of the Friedberg and Babbitt labs for
insightful discussions. We are especially grateful to Rachael Huntley for careful reading of the
manuscript and for providing 
detailed explanations of annotation provenance and methodology in UniProt.

\section*{Funding}
This research was funded, in part, by 
NSF DBI 1146960 award to IF and 
NIH R01 GM60595 and NSF DBI
0640476 to PCB.

\bibliography{sp_bias.bib}
\newpage
\section*{Figures}

\newpage

\newpage

\newpage
\begin{figure}[!ht]
\begin{center}
\includegraphics[width=6.5in]{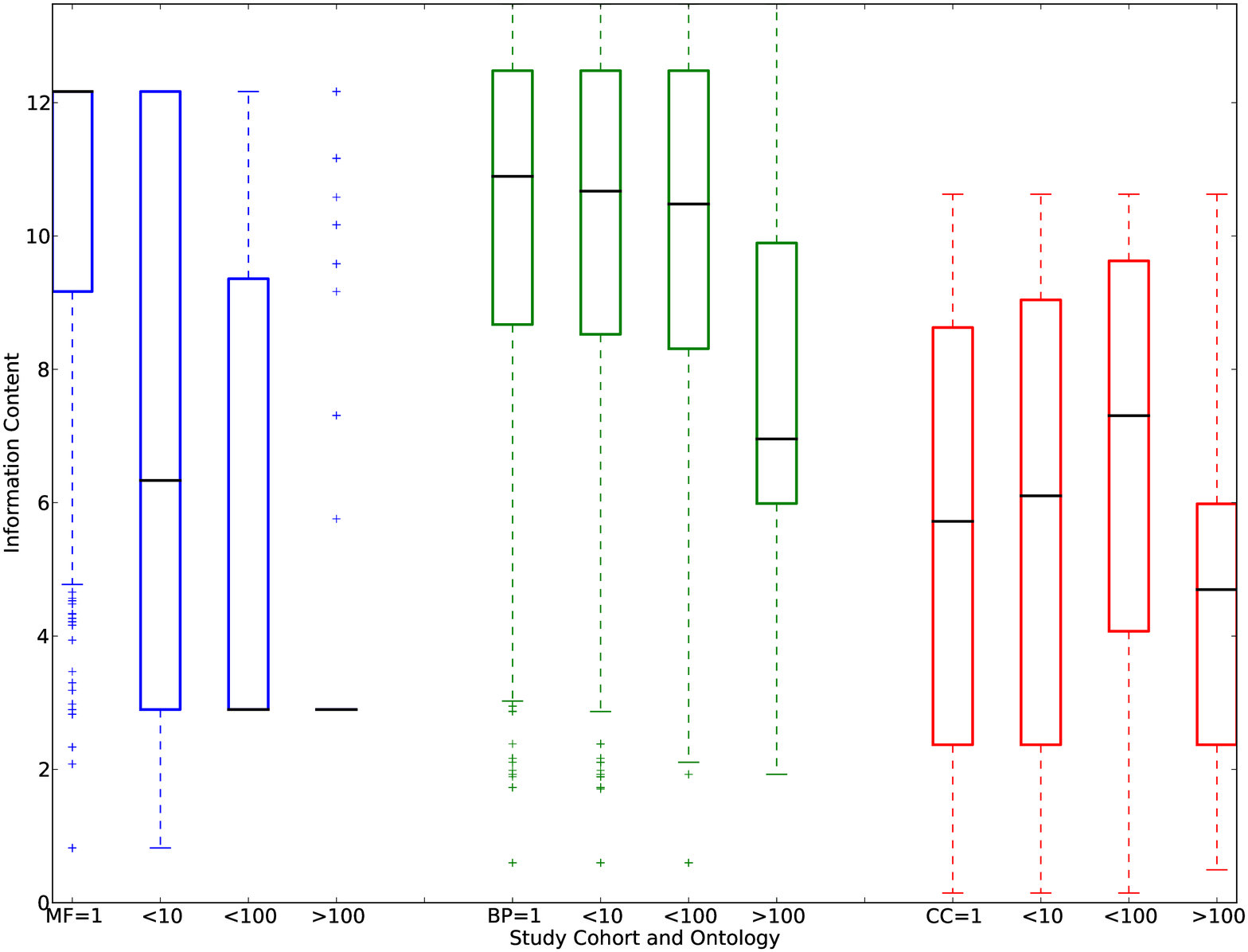}
\end{center}
\end{figure}
\newpage
\textbf{4B}\\
\newpage
\clearpage

\section*{Tables}

\begin{table}[!ht]
\caption{
\bf{Annotation Cohorts}}
\begin{tabular}{||p{5cm}||l|l|l|l||l||}
\hline
\textbf{Articles annotating the following number of proteins} & 
\textbf{1} & \textbf{$1<n<10$} & \textbf{$10\leq n<100$} & \textbf{$n\geq100$} 
& \textbf{SUM} \\ \hline
\textbf{Number of proteins annotated} & 20699 & 46383 & 26485 & 31411 & 124978 \\ \hline
\textbf{Number of annotating articles} & 41156 & 32201 & 2672 & 108 &  76137 \\ \hline
\textbf{Percent of proteins annotated} & 16.56 & 37.11 & 21.19 & 25.13 & 100 \\ \hline
\textbf{Percent of annotating articles} & 54.09 & 42.32 & 3.51 & 0.14 & 100 \\ \hline 
\end{tabular}
\begin{flushleft} Number of proteins and annotating articles assigned to each article annotation
cohort. Columns: 1: articles annotating a single protein (singletons); 
$1<n<10$ articles annotating more than 1 and less than 10 proteins (low throughput);
$10\leq n<100$:  medium throughput;
$n\geq100$: articles annotating 100 proteins and more (high throughput).
As can be seen, high-throughput articles comprise 0.14\% of the total articles used for experimental
annotations, but annotate 25.13\% of the proteins in UniProt-GOA.
\end{flushleft}
\label{tab:cohorts}
\end{table}
\newpage

\begin{table}[!ht]
\caption{
\bf{Sequence Redundancy in Top-50 Annotating Articles}}
\begin{tabular}{|p{3cm}|p{1.5cm}|p{2cm}|p{2cm}|p{2cm}|p{2cm}|} 
\hline
\textbf{Species} & \textbf{num. articles} &\textbf{num. prot} & \textbf{Clusters at 100\%} & 
\textbf{\% redundancy} & \textbf{Mean genes/ cluster} 
\\ \hline
\textit{C. elegans}  & 12 & 8416 & 3338 & 60 & 3.74 
\\ \hline
\textit{A. thaliana}  & 16 &  8879 & 4694 & 47 & 3.92 
\\ \hline
\textit{M. musculus}  & 3 &  4220 & 2273 & 46 & 2.75  
\\ \hline
\textit{M. tuberculosis}  & 2 & 2351& 1702& 28 & 2.22 
\\ \hline
\textit{S. cerevisiae}  & 5 & 3542 & 2550 & 28 & 2.33 
\\ \hline
\textit{H. sapiens}  & 4 & 5593 & 4509 & 19 & 2.36 
\\ \hline
\textit{D. melanogaster} & 3  & 1217 & 1003 & 18 & 2.17 
\\ \hline
\textit{S. pombe}  & 2 & 4502 & 4281 & 5 & 2.00 
\\ \hline
\end{tabular}
\begin{flushleft} 
\textbf{Species}: annotated species; 
\textbf{num. articles} number of annotating articles;
\textbf{num. prot}: number of proteins annotated by top-50 articles for that species; 
\textbf{Clusters at 100\%}: number of clusters of 100\% identical proteins; 
\textbf{\% redundancy}: the product of  column 4 by column 3: this is the percentage of proteins
annotated more than once for a given species in the top 50 articles; 
\textbf{Mean genes/cluster}: the mean number of genes per cluster, for clusters
having more than a single gene.
\end{flushleft}
\label{tab:dreamcatcher1}
\end{table}
\newpage

\begin{table}[!ht]
\caption{
\bf{Annotation Consistency in Top 50 articles}}
\begin{tabular}{|p{3cm}|l|l|l|l|l|p{1.5cm}|p{1.5cm}|}
\hline
\textbf{Species} & \textbf{Ont.} & \textbf{num prot} & \textbf{mean} 
$k_{P,O}$ &
\textbf{stdv} & \textbf{stderr} & \textbf{num articles} & \textbf{num terms} \\
\hline \hline
A. thaliana & CCO & 1941 & 0.251 & 0.328 & 0.007 & 15 & 18\\ \hline
C. elegans & BPO & 1847 & 0.388 & 0.239 & 0.006 & 12 & 41\\ \hline 
D. melanogaster & BPO & 76 & 0.086 & 0.22 & 0.025 & 3 & 8\\ \hline
D. melanogaster & CCO & 81 & 0.068 & 0.234 & 0.026 & 3 & 5\\ \hline
H. sapiens & CCO & 167 & 0.285 & 0.365 & 0.028 & 2 & 20\\ \hline
M. musculus & CCO & 807 & 0.832 & 0.291 & 0.01 & 3 & 2\\ \hline
S. cerevisiae & CCO & 744 & 0.759 & 0.379 & 0.014 & 4 & 15\\ \hline
B. tuberculosis & CCO & 532 & 0.309 & 0.41 & 0.018 & 2 & 3\\ \hline
\end{tabular}
\begin{flushleft}
\textbf{Species}: annotated species;
\textbf{Ontology}: annotating GO ontology;
\textbf{num prot}: number of annotated proteins in that species \& ontology that are annotated by
more than one paper.
\textbf{mean, stdv, stderr}: mean number of consistent annotations for a protein in that species
and ontology, standard deviation from the mean and standard error.
\textbf{num articles}: number of annotating articles
\textbf{num terms} number of annotating terms. 
Annotations by less than two articles or two terms (or both) for the same protein/ontology
combination have been omitted.
\end{flushleft}
\label{tab:dreamcatcher2}
\end{table}
\newpage
\begin{table}[!ht]
\caption{
\bf{Fraction of Proteins Exclusively Annotated by High Throughput Studies}} 
\begin{tabular}{|l|l|l|l|l|} 
\hline
\textbf{Taxon ID} & \textbf{Taxon} & \textbf{XHT} & 
\textbf{Total Proteins} & \textbf{\%XHT} \\
\hline
\hline
284812 & Schizosaccharomyces pombe & 2781 & 4507 & 61.704 \\ 
\hline
1773 & Bacillus tuberculosis & 1224 & 2317 & 52.8269 \\ 
\hline
6239 & Caenorhabditis elegans & 2493 & 5302 & 47.02 \\ 
\hline
9606 & Homo sapiens & 4016 & 11521 & 34.8581 \\ 
\hline
44689 & Dictyostelium discoideum & 425 & 1256 & 33.8376 \\ 
\hline
3702 & Arabidopsis thaliana & 3199 & 10153 & 31.5079 \\ 
\hline
237561 & Candida albicans SC5314 & 327 & 1243 & 26.3073 \\ 
\hline
10090 & LK3 transgenic mice & 2567 & 22068 & 11.6322 \\ 
\hline
7227 & Drosophila melanogaster & 735 & 7501 & 9.7987 \\ 
\hline
559292 & Saccharomyces cerevisiae & 439 & 5086 & 8.6315 \\ 
\hline
83333 & Escherichia coli K-12 & 83 & 1606 & 5.1681 \\ 
\hline
7955 & Brachidanio rerio & 117 & 4633 & 2.5254 \\ 
\hline
10116 & Buffalo rat & 11 & 4634 & 0.2374 \\ 
\hline
\end{tabular}
\begin{flushleft}
\textbf{Taxon ID}: NCBI Taxon ID number;
\textbf{Species}: annotated species;
\textbf{XHT}: number of proteins exclusively annotated by high-throughput experimental
studies (100 or more proteins annotated per study);
\textbf{Total proteins}: Total number of experimentally annotated proteins in that species.
\textbf{\%XHT}: percentage of proteins in that species that are annotated exclusively by HT
studies.
\end{flushleft}
\label{tab:exclusive_ht}
\end{table}

\newpage
\section*{Supplementary Material Legends}

\noindent
\textbf{Table \ref{tab:top50}:
The top 50 annotating articles.} \\
\textbf{N}: article rank; \textbf{Proteins}: number of proteins annotated in this article;
\textbf{Annotations}: number of annotating GO terms; \textbf{Species}: annotated species;
\textbf{ref.} annotating article; \textbf{MFO}/\textbf{BPO}/\textbf{CCO}: number of
proteins annotated in the Molecular Function, Biological Process and Cellular Component
ontologies, respectively.
\\\\
\noindent
\textbf{Table \ref{tab:sup:papers_eco}:
The Top-50 studies and the ECO terms we have assigned to them.} \\
\textbf{PMID}: Articles' PubMed ID; \textbf{ECO terms/ECO ID's}: terms
and ID's we assigned to the articles. 
\\\\
\noindent
\textbf{Table \ref{tab:evidence}:
ECO terms were assigned by us to the top-50 annotating papers.} \\
The table
entries are ranked by the frequency of the assignments, i.e. 27 papers
are assigned with term ECO:0000160, 21 were assigned ECO:0000004, etc.
Entries in \textbf{boldface} are for computational methods, which were
used in many papers in combination with experimental methods to assign
function. Table \ref{tab:sup:papers_eco} lists the ECO terms. 
\newpage
\startsupplement

\begin{longtable}[!ht]{|l|l|l|l|l|l|l|l|}
\caption{\textbf{Top 50 Annotating Articles}} \\
\hline
\textbf{N}&\textbf{Proteins}&\textbf{Annotations}&\textbf{Species}&
\textbf{ref.}&
\textbf{MFO}& \textbf{BPO}&\textbf{CCO}\\\hline
\endfirsthead
\hline
\textbf{N}&\textbf{Proteins}&\textbf{Annotations}&\textbf{Species}&
\textbf{ref.} &
\textbf{MFO}&\textbf{BPO}&\textbf{CCO} \\\hline
\endhead
\hline \multicolumn{4}{r}{\textit{Continued on next page}} \\
\endfoot
\hline
\endlastfoot
1 & 4937 & 11050 & \textit{H. sapiens} & \cite{pmid18029348} & 0 & 0 & 11050\\ \hline
2 & 4247 & 7046 & \textit{S. pombe} & \cite{pmid16823372} & 0 & 0 & 7046\\ \hline
3 & 2412 & 2412 & \textit{H. sapiens} & \cite{pmid18614015} & 0 & 0 & 2412\\ \hline
4 & 1791 & 5918 & \textit{C. elegans} & \cite{pmid14551910} & 0 & 5918 & 0\\ \hline
5 & 1406 & 1863 & \textit{S. cerevisiae} & \cite{pmid14562095} & 0 & 0 & 1863\\ \hline
6 & 1251 & 1251 & \textit{A. thaliana} & \cite{pmid18431481} & 0 & 0 & 1251\\ \hline
7 & 1205 & 1476 & \textit{C. elegans} & \cite{pmid15791247} & 0 & 1476 & 0\\ \hline
8 & 1186 & 1213 & \textit{M. musculus} & \cite{pmid14651853} & 0 & 0 & 1213\\ \hline
9 & 1136 & 1136 & \textit{A. thaliana} & \cite{pmid17317660} & 0 & 0 & 1136\\ \hline
10 & 1101 & 2269 & \textit{C. elegans} & \cite{pmid12529635} & 0 & 2269 & 0\\ \hline
11 & 1043 & 1365 & \textit{M. tuberculosis} & \cite{pmid15525680} & 0 & 0 & 1365\\ \hline
12 & 1041 & 1041 & \textit{A. thaliana} & \cite{pmid21166475} & 0 & 0 & 1041\\ \hline
13 & 865 & 1533 & \textit{C. elegans} & \cite{pmid15489339} & 0 & 1533 & 0\\ \hline
14 & 845 & 845 & \textit{S. cerevisiae} & \cite{pmid16823961} & 0 & 0 & 845\\ \hline
15 & 784 & 784 & \textit{A. thaliana} & \cite{pmid21533090} & 0 & 0 & 784\\ \hline
16 & 735 & 735 & \textit{M. tuberculosis} & \cite{pmid14532352} & 0 & 0 & 735\\ \hline
17 & 724 & 882 & \textit{A. thaliana} & \cite{pmid20061580} & 0 & 0 & 882\\ \hline
18 & 634 & 634 & \textit{A. thaliana} & \cite{pmid15028209} & 0 & 0 & 634\\ \hline
19 & 613 & 613 & Mycobacter sp. & \cite{pmid12657046} & 0 & 613 & 0\\ \hline
20 & 607 & 661 & \textit{C. elegans} & \cite{pmid17704769} & 0 & 659 & 2\\ \hline
21 & 577 & 577 & \textit{A. thaliana} & \cite{pmid17432890} & 0 & 0 & 577\\ \hline
22 & 553 & 884 & \textit{C. elegans} & \cite{pmid11231151} & 0 & 884 & 0\\ \hline
23 & 516 & 5972 & \textit{C. elegans} & \cite{pmid17417969} & 0 & 5972 & 0\\ \hline
24 & 503 & 503 & \textit{S. cerevisiae} & \cite{pmid14576278} & 0 & 0 & 503\\ \hline
25 & 498 & 638 & \textit{S. cerevisiae} & \cite{pmid16429126} & 638 & 0 & 0\\ \hline
26 & 479 & 848 & \textit{C. elegans} & \cite{pmid21529718} & 0 & 848 & 0\\ \hline
27 & 465 & 468 & \textit{H. sapiens} & \cite{pmid11256614} & 0 & 0 & 468\\ \hline
28 & 436 & 436 & \textit{A. thaliana} & \cite{pmid17644812} & 0 & 0 & 436\\ \hline
29 & 430 & 513 & \textit{A. thaliana} & \cite{pmid16618929} & 0 & 0 & 513\\ \hline
30 & 413 & 456 &  \textit{D. melanogaster} & \cite{pmid18433294} & 0 & 39 & 417\\ \hline
31 & 401 & 401 & \textit{A. thaliana} & \cite{pmid17151019} & 0 & 0 & 401\\ \hline
32 & 392 & 392 & \textit{A. thaliana} & \cite{pmid14671022} & 0 & 0 & 392\\ \hline
33 & 392 & 639 & \textit{C. elegans} & \cite{pmid12529643} & 0 & 639 & 0\\ \hline
34 & 383 & 917 & \textit{C. elegans} & \cite{pmid12445391} & 0 & 917 & 0\\ \hline
35 & 380 & 380 & \textit{A. thaliana} & \cite{pmid15539469} & 0 & 0 & 380\\ \hline
36 & 375 & 375 & \textit{M. musculus} & \cite{pmid12865426} & 0 & 0 & 375\\ \hline
37 & 343 & 509 & \textit{H. sapiens} & \cite{pmid16189514} & 509 & 0 & 0\\ \hline
38 & 338 & 338 & Ddiscoideum & \cite{pmid20422638} & 0 & 0 & 338\\ \hline
39 & 328 & 328 & \textit{A. thaliana} & \cite{pmid12938931} & 0 & 0 & 328\\ \hline
40 & 319 & 329 & \textit{C. albicans} & \cite{pmid16336044} & 1 & 328 & 0\\ \hline
41 & 305 & 312 & \textit{A. thaliana} & \cite{pmid18633119} & 0 & 0 & 312\\ \hline
42 & 290 & 331 & \textit{S. cerevisiae} & \cite{pmid11914276} & 0 & 0 & 331\\ \hline
43 & 285 & 761 & \textit{C. elegans} & \cite{pmid11099033} & 0 & 761 & 0\\ \hline
44 & 283 & 499 & \textit{C. elegans} & \cite{pmid11099034} & 0 & 499 & 0\\ \hline
45 & 266 & 433 & \textit{M. musculus} & \cite{pmid11591653} & 433 & 0 & 0\\ \hline
46 & 260 & 260 & \textit{A. thaliana} & \cite{pmid16502469} & 0 & 260 & 0\\ \hline
47 & 258 & 259 & \textit{S. pombe} & \cite{pmid12529438} & 0 & 259 & 0\\ \hline
48 & 244 & 397 &  \textit{D. melanogaster} & \cite{pmid17412918} & 0 & 367 & 30\\ \hline
49 & 242 & 397 &  \textit{D. melanogaster} & \cite{pmid18981222} & 0 & 0 & 397\\ \hline
50 & 241 & 263 & \textit{A. thaliana} & \cite{pmid16287169} & 0 & 0 & 263\\ \hline
\label{tab:top50}
\end{longtable}
\begin{flushleft} {\bf The top 50 annotating articles.} \textbf{N}:
article rank; \textbf{Proteins}: number of proteins annotated in this
article; \textbf{Annotations}: number of annotating
GO terms; \textbf{Species}: annotated species; \textbf{ref.} annotating
article; \textbf{MFO}/\textbf{BPO}/\textbf{CCO}: number of proteins
annotated in the Molecular Function, Biological Process and Cellular
Component ontologies, respectively.
\end{flushleft}

\newpage

%
\begin{longtable}[h]{|l|l|p{12cm}|}
\caption{\textbf{ECO Terms Assigned to Top-50 Papers}} \\
\hline
\textbf{PMID} & \textbf{Ref} & \textbf{ECO terms/ECO ID's} \\
\endfirsthead
\hline
\textbf{PMID} & \textbf{Ref} & \textbf{ECO terms/ECO ID's} \\
\endhead
\hline
\multicolumn{3}{r}{\textit{Continued on next page}} \\
\endfoot
\hline
\endlastfoot
18029348 & \cite{pmid18029348} & imaging assay evidence/ECO:0000324   immunofluorescence evidence/ECO:0000007   immunolocalization evidence/ECO:0000087 \\ 
\hline
16823372 & \cite{pmid16823372} & imaging assay evidence/ECO:0000324   yellow fluorescent protein fusion protein localization evidence/ECO:0000128   enzyme inhibition experiment evidence/ECO:0000184 \\ 
\hline
18614015 & \cite{pmid18614015} & imaging assay evidence/ECO:0000324   protein separation followed by fragment identification evidence/ECO:0000160   sequence similarity evidence/ECO:0000044   cell fractionation evidence/ECO:0000004   GFP fusion protein localization evidence/ECO:0000126   computational combinatorial evidence/ECO:0000053   motif similarity evidence/ECO:0000028   targeting sequence prediction evidence/ECO:0000081   protein BLAST evidence/ECO:0000208 \\ 
\hline
14551910 & \cite{pmid14551910} & imaging assay evidence/ECO:0000324   RNAi evidence/ECO:0000019   loss-of-function mutant phenotype evidence/ECO:0000016   nucleotide BLAST evidence/ECO:0000207   sequence alignment evidence/ECO:0000200 \\ 
\hline
14562095 & \cite{pmid14562095} & imaging assay evidence/ECO:0000324   GFP fusion protein localization evidence/ECO:0000126   fusion protein localization evidence/ECO:0000124   affinity chromatography evidence/ECO:0000079 \\ 
\hline
18431481 & \cite{pmid18431481} & protein separation followed by fragment identification evidence/ECO:0000160   targeting sequence prediction evidence/ECO:0000081   cell fractionation evidence/ECO:0000004   sequence similarity evidence/ECO:0000044   imported information/ECO:0000311 \\ 
\hline
15791247 & \cite{pmid15791247} & imaging assay evidence/ECO:0000324   RNAi evidence/ECO:0000019   loss-of-function mutant phenotype evidence/ECO:0000016   protein BLAST evidence/ECO:0000208 \\ 
\hline
14651853 & \cite{pmid14651853} & protein separation followed by fragment identification evidence/ECO:0000160   cell fractionation evidence/ECO:0000004   targeting sequence prediction evidence/ECO:0000081   sequence similarity evidence/ECO:0000044   protein BLAST evidence/ECO:0000208   nucleotide BLAST evidence/ECO:0000207   Affymetrix array experiment evidence/ECO:0000101   imported information/ECO:0000311 \\ 
\hline
17317660 & \cite{pmid17317660} & protein separation followed by fragment identification evidence/ECO:0000160   cell fractionation evidence/ECO:0000004   transmembrane domain prediction evidence/ECO:0000083   sequence similarity evidence/ECO:0000044 \\ 
\hline
12529635 & \cite{pmid12529635} & imaging assay evidence/ECO:0000324   RNAi evidence/ECO:0000019   loss-of-function mutant phenotype evidence/ECO:0000016   motif similarity evidence/ECO:0000028   protein BLAST evidence/ECO:0000208   nucleotide BLAST evidence/ECO:0000207   computational combinatorial evidence/ECO:0000053 \\ 
\hline
15525680 & \cite{pmid15525680} & protein separation followed by fragment identification evidence/ECO:0000160   cell fractionation evidence/ECO:0000004   transmembrane domain prediction evidence/ECO:0000083   sequence similarity evidence/ECO:0000044   computational combinatorial evidence/ECO:0000053   biological system reconstruction/ECO:0000088   imported information/ECO:0000311   protein BLAST evidence/ECO:0000208 \\ 
\hline
21166475 & \cite{pmid21166475} & protein separation followed by fragment identification evidence/ECO:0000160   cell fractionation evidence/ECO:0000004   sequence similarity evidence/ECO:0000044   computational combinatorial evidence/ECO:0000053   imported information/ECO:0000311   transmembrane domain prediction evidence/ECO:0000083   sequence alignment evidence/ECO:0000200   motif similarity evidence/ECO:0000028 \\ 
\hline
15489339 & \cite{pmid15489339} & imaging assay evidence/ECO:0000324   RNAi evidence/ECO:0000019   loss-of-function mutant phenotype evidence/ECO:0000016   nucleotide BLAST evidence/ECO:0000207 \\ 
\hline
16823961 & \cite{pmid16823961} & protein separation followed by fragment identification evidence/ECO:0000160   cell fractionation evidence/ECO:0000004   sequence similarity evidence/ECO:0000044   imported information/ECO:0000311 \\ 
\hline
21533090 & \cite{pmid21533090} & protein separation followed by fragment identification evidence/ECO:0000160   cell fractionation evidence/ECO:0000004   sequence similarity evidence/ECO:0000044   imported information/ECO:0000311   computational combinatorial evidence/ECO:0000053   transmembrane domain prediction evidence/ECO:0000083   sequence alignment evidence/ECO:0000200   motif similarity evidence/ECO:0000028   targeting sequence prediction evidence/ECO:0000081 \\ 
\hline
14532352 & \cite{pmid14532352} & protein separation followed by fragment identification evidence/ECO:0000160   cell fractionation evidence/ECO:0000004   sequence similarity evidence/ECO:0000044   transmembrane domain prediction evidence/ECO:0000083 \\ 
\hline
20061580 & \cite{pmid20061580} & protein separation followed by fragment identification evidence/ECO:0000160   cell fractionation evidence/ECO:0000004   sequence similarity evidence/ECO:0000044   transmembrane domain prediction evidence/ECO:0000083   imported information/ECO:0000311   targeting sequence prediction evidence/ECO:0000081   protein expression level evidence/ECO:0000046 \\ 
\hline
15028209 & \cite{pmid15028209} & protein separation followed by fragment identification evidence/ECO:0000160   cell fractionation evidence/ECO:0000004   sequence similarity evidence/ECO:0000044   targeting sequence prediction evidence/ECO:0000081   Affymetrix array experiment evidence/ECO:0000101   protein expression level evidence/ECO:0000046   protein BLAST evidence/ECO:0000208   computational combinatorial evidence/ECO:0000053   motif similarity evidence/ECO:0000028   transmembrane domain prediction evidence/ECO:0000083 \\ 
\hline
12657046 & \cite{pmid12657046} & mutant phenotype evidence/ECO:0000015   nucleic acid hybridization evidence/ECO:0000026   imported information/ECO:0000311   sequence similarity evidence/ECO:0000044   combinatorial evidence/ECO:0000212 \\ 
\hline
17704769 & \cite{pmid17704769} & imaging assay evidence/ECO:0000324   RNAi evidence/ECO:0000019   loss-of-function mutant phenotype evidence/ECO:0000016 \\ 
\hline
17432890 & \cite{pmid17432890} & protein separation followed by fragment identification evidence/ECO:0000160   cell fractionation evidence/ECO:0000004   sequence similarity evidence/ECO:0000044   transmembrane domain prediction evidence/ECO:0000083   imported information/ECO:0000311   targeting sequence prediction evidence/ECO:0000081   protein BLAST evidence/ECO:0000208   computational combinatorial evidence/ECO:0000053 \\ 
\hline
11231151 & \cite{pmid11231151} & imaging assay evidence/ECO:0000324   RNAi evidence/ECO:0000019   loss-of-function mutant phenotype evidence/ECO:0000016 \\ 
\hline
17417969 & \cite{pmid17417969} & imaging assay evidence/ECO:0000324   RNAi evidence/ECO:0000019   loss-of-function mutant phenotype evidence/ECO:0000016 \\ 
\hline
14576278 & \cite{pmid14576278} & protein separation followed by fragment identification evidence/ECO:0000160   cell fractionation evidence/ECO:0000004   sequence similarity evidence/ECO:0000044   transmembrane domain prediction evidence/ECO:0000083 \\ 
\hline
16429126 & \cite{pmid16429126} & protein separation followed by fragment identification evidence/ECO:0000160   sequence similarity evidence/ECO:0000044   affinity chromatography evidence/ECO:0000079   protein BLAST evidence/ECO:0000208   imported information/ECO:0000311 \\ 
\hline
21529718 & \cite{pmid21529718} & imaging assay evidence/ECO:0000324   RNAi evidence/ECO:0000019   loss-of-function mutant phenotype evidence/ECO:0000016   computational combinatorial evidence/ECO:0000053 \\ 
\hline
11256614 & \cite{pmid11256614} & GFP fusion protein localization evidence/ECO:0000126   yellow fluorescent protein fusion protein localization evidence/ECO:0000128   imaging assay evidence/ECO:0000324   motif similarity evidence/ECO:0000028   protein BLAST evidence/ECO:0000208   nucleotide BLAST evidence/ECO:0000207 \\ 
\hline
17644812 & \cite{pmid17644812} & protein separation followed by fragment identification evidence/ECO:0000160   cell fractionation evidence/ECO:0000004   sequence similarity evidence/ECO:0000044   transmembrane domain prediction evidence/ECO:0000083   targeting sequence prediction evidence/ECO:0000081   computational combinatorial evidence/ECO:0000053 \\ 
\hline
16618929 & \cite{pmid16618929} & protein separation followed by fragment identification evidence/ECO:0000160   cell fractionation evidence/ECO:0000004   sequence similarity evidence/ECO:0000044   transmembrane domain prediction evidence/ECO:0000083 \\ 
\hline
18433294 & \cite{pmid18433294} & protein separation followed by fragment identification evidence/ECO:0000160   cell fractionation evidence/ECO:0000004   sequence similarity evidence/ECO:0000044   imaging assay evidence/ECO:0000324   RNAi evidence/ECO:0000019   loss-of-function mutant phenotype evidence/ECO:0000016   immunofluorescence evidence/ECO:0000007 \\ 
\hline
17151019 & \cite{pmid17151019} & protein separation followed by fragment identification evidence/ECO:0000160   cell fractionation evidence/ECO:0000004   sequence similarity evidence/ECO:0000044   imported information/ECO:0000311   transmembrane domain prediction evidence/ECO:0000083 \\ 
\hline
14671022 & \cite{pmid14671022} & protein separation followed by fragment identification evidence/ECO:0000160   cell fractionation evidence/ECO:0000004   sequence similarity evidence/ECO:0000044   protein BLAST evidence/ECO:0000208   targeting sequence prediction evidence/ECO:0000081 \\ 
\hline
12529643 & \cite{pmid12529643} & imaging assay evidence/ECO:0000324   RNAi evidence/ECO:0000019   loss-of-function mutant phenotype evidence/ECO:0000016 \\ 
\hline
12445391 & \cite{pmid12445391} & imaging assay evidence/ECO:0000324   RNAi evidence/ECO:0000019   loss-of-function mutant phenotype evidence/ECO:0000016   BLAST evidence/ECO:0000206 \\ 
\hline
15539469 & \cite{pmid15539469} & protein separation followed by fragment identification evidence/ECO:0000160   cell fractionation evidence/ECO:0000004   sequence similarity evidence/ECO:0000044   targeting sequence prediction evidence/ECO:0000081   transmembrane domain prediction evidence/ECO:0000083   motif similarity evidence/ECO:0000028   protein BLAST evidence/ECO:0000208   computational combinatorial evidence/ECO:0000053 \\ 
\hline
12865426 & \cite{pmid12865426} & protein separation followed by fragment identification evidence/ECO:0000160   cell fractionation evidence/ECO:0000004   sequence similarity evidence/ECO:0000044   transmembrane domain prediction evidence/ECO:0000083 \\ 
\hline
16189514 & \cite{pmid16189514} & yeast 2-hybrid evidence/ECO:0000068   imaging assay evidence/ECO:0000324   motif similarity evidence/ECO:0000028   co-purification evidence/ECO:0000022   combinatorial evidence/ECO:0000212 \\ 
\hline
20422638 & \cite{pmid20422638} & protein separation followed by fragment identification evidence/ECO:0000160   cell fractionation evidence/ECO:0000004   sequence similarity evidence/ECO:0000044   combinatorial evidence/ECO:0000212 \\ 
\hline
12938931 & \cite{pmid12938931} & protein separation followed by fragment identification evidence/ECO:0000160   cell fractionation evidence/ECO:0000004   sequence similarity evidence/ECO:0000044   nucleotide BLAST evidence/ECO:0000207   imported information/ECO:0000311   transmembrane domain prediction evidence/ECO:0000083 \\ 
\hline
16336044 & \cite{pmid16336044} & imaging assay evidence/ECO:0000324   RNAi evidence/ECO:0000019   loss-of-function mutant phenotype evidence/ECO:0000016 \\ 
\hline
18633119 & \cite{pmid18633119} & protein separation followed by fragment identification evidence/ECO:0000160   cell fractionation evidence/ECO:0000004   Western blot evidence/ECO:0000112 \\ 
\hline
11914276 & \cite{pmid11914276} & imaging assay evidence/ECO:0000324   immunofluorescence evidence/ECO:0000007   epitope-tagged protein immunolocalization evidence/ECO:0000092   transmembrane domain prediction evidence/ECO:0000083   imported information/ECO:0000311 \\ 
\hline
11099033 & \cite{pmid11099033} & imaging assay evidence/ECO:0000324   RNAi evidence/ECO:0000019   loss-of-function mutant phenotype evidence/ECO:0000016   protein BLAST evidence/ECO:0000208   computational combinatorial evidence/ECO:0000053 \\ 
\hline
11099034 & \cite{pmid11099034} & imaging assay evidence/ECO:0000324   RNAi evidence/ECO:0000019   loss-of-function mutant phenotype evidence/ECO:0000016   nucleotide BLAST evidence/ECO:0000207   protein BLAST evidence/ECO:0000208 \\ 
\hline
11591653 & \cite{pmid11591653} & hybrid interaction evidence/ECO:0000025   imaging assay evidence/ECO:0000324 \\ 
\hline
16502469 & \cite{pmid16502469} & protein separation followed by fragment identification evidence/ECO:0000160   sequence similarity evidence/ECO:0000044   protein BLAST evidence/ECO:0000208   Northern assay evidence/ECO:0000106   reverse transcription polymerase chain reaction transcription evidence/ECO:0000108 \\ 
\hline
12529438 & \cite{pmid12529438} & microarray RNA expression level evidence/ECO:0000104   sequence orthology evidence used in manual assertion/ECO:0000266   motif similarity evidence/ECO:0000028 \\ 
\hline
17412918 & \cite{pmid17412918} & RNAi evidence/ECO:0000019   loss-of-function mutant phenotype evidence/ECO:0000016   imaging assay evidence/ECO:0000324 \\ 
\hline
18981222 & \cite{pmid18981222} & protein separation followed by fragment identification evidence/ECO:0000160   sequence similarity evidence/ECO:0000044   protein BLAST evidence/ECO:0000208   in vitro assay evidence/ECO:0000181   affinity chromatography evidence/ECO:0000079   imaging assay evidence/ECO:0000324   mutant phenotype evidence/ECO:0000015 \\ 
\hline
16287169 & \cite{pmid16287169} & protein separation followed by fragment identification evidence/ECO:0000160   sequence similarity evidence/ECO:0000044   transmembrane domain prediction evidence/ECO:0000083   sequence alignment evidence/ECO:0000200   computational combinatorial evidence/ECO:0000053   motif similarity evidence/ECO:0000028   targeting sequence prediction evidence/ECO:0000081 
\label{tab:sup:papers_eco}
\end{longtable}
\begin{flushleft}

The Top-50 studies and the ECO terms we have assigned to them. \textbf{PMID}: Articles' PubMed ID;
\textbf{ECO terms/ECO ID's}: terms and ID's we assigned to the articles.
\end{flushleft}

\newpage

%
\begin{longtable}[h]{|l|p{8cm}|l|l|}
\caption{\textbf{Count of ECO terms in top-50 papers}} \\
\hline
\textbf{N} & \textbf{ECO term} & \textbf{ECO ID} & \textbf{Articles} \\ \hline
\endfirsthead
\hline
\textbf{N} & \textbf{ECO term} & \textbf{ECO ID} & \textbf{Articles} \\ \hline
\endhead
\hline
\multicolumn{4}{r}{\textit{Continued on next page}} \\
\endfoot
\hline
\endlastfoot
1 & protein separation followed by fragment identification evidence & ECO:0000160 & 27 \\ 
\hline
2 & \textbf{sequence similarity evidence} & ECO:0000044 & 27 \\ 
\hline
3 & imaging assay evidence & ECO:0000324 & 24 \\ 
\hline
4 & cell fractionation evidence & ECO:0000004 & 23 \\ 
\hline
5 & \textbf{transmembrane domain prediction evidence} & ECO:0000083 & 17 \\ 
\hline
6 & loss-of-function mutant phenotype evidence & ECO:0000016 & 15 \\ 
\hline
7 & \textbf{protein BLAST evidence} & ECO:0000208 & 15 \\ 
\hline
8 & RNAi evidence & ECO:0000019 & 15 \\ 
\hline
9 & imported information & ECO:0000311 & 13 \\ 
\hline
10 & \textbf{computational combinatorial evidence} & ECO:0000053 & 11 \\ 
\hline
11 & \textbf{targeting sequence prediction evidence} & ECO:0000081 & 11 \\ 
\hline
12 & \textbf{motif similarity evidence} & ECO:0000028 & 10 \\ 
\hline
13 & \textbf{nucleotide BLAST evidence} & ECO:0000207 & 7 \\ 
\hline
14 & \textbf{sequence alignment evidence} & ECO:0000200 & 4 \\ 
\hline
15 & GFP fusion protein localization evidence & ECO:0000126 & 3 \\ 
\hline
16 & immunofluorescence evidence & ECO:0000007 & 3 \\ 
\hline
17 & affinity chromatography evidence & ECO:0000079 & 3 \\ 
\hline
18 & \textbf{computational combinatorial evidence} & ECO:0000053 & 2 \\ 
\hline
19 & Affymetrix array experiment evidence & ECO:0000101 & 2 \\ 
\hline
20 & protein expression level evidence & ECO:0000046 & 2 \\ 
\hline
21 & mutant phenotype evidence & ECO:0000015 & 2 \\ 
\hline
22 & combinatorial evidence & ECO:0000212 & 2 \\ 
\hline
23 & co-purification evidence & ECO:0000022 & 1 \\ 
\hline
24 & epitope-tagged protein immunolocalization evidence & ECO:0000092 & 1 \\ 
\hline
25 & \textbf{sequence orthology evidence used in manual assertion} & ECO:0000266 & 1 \\ 
\hline
26 & YFP fusion protein localization evidence & ECO:0000128 & 2 \\ 
\hline
27 & in vitro assay evidence & ECO:0000181 & 1 \\ 
\hline
28 & biological system reconstruction & ECO:0000088 & 1 \\ 
\hline
29 & reverse transcription polymerase chain reaction transcription evidence & ECO:0000108 & 1 \\ 
\hline
30 & Northern assay evidence & ECO:0000106 & 1 \\ 
\hline
31 & Western blot evidence & ECO:0000112 & 1 \\ 
\hline
32 & microarray RNA expression level evidence & ECO:0000104 & 1 \\ 
\hline
33 & fusion protein localization evidence & ECO:0000124 & 1 \\ 
\hline
34 & \textbf{BLAST evidence} & ECO:0000206 & 1 \\ 
\hline
35 & nucleic acid hybridization evidence & ECO:0000026 & 1 \\ 
\hline
36 & enzyme inhibition experiment evidence & ECO:0000184 & 1 \\ 
\hline
37 & immunolocalization evidence & ECO:0000087 & 1 \\ 
\hline
38 & hybrid interaction evidence & ECO:0000025 & 1 \\ 
\hline
39 & yeast 2-hybrid evidence & ECO:0000068 & 1 
\label{tab:evidence}
\end{longtable}
\begin{flushleft}
ECO terms were assigned by us to the top-50 annotating papers. The table entries are ranked
by the frequency of the assignments, i.e. 27 papers are assigned with term ECO:0000160, 21
were assigned ECO:0000004, etc. Entries in \textbf{boldface} are for computational methods,
which were used in many papers in combination with experimental methods to assign function. 
Table\ref{tab:sup:papers_eco} lists the ECO terms. 

\end{flushleft}

\newpage

\end{document}